\documentclass[aps,prl,twocolumn,longbibliography,epsfig,floats,preprintnumbers,floatfix]{revtex4-2}
\usepackage{graphicx,amsmath,amssymb,bm,color,sidecap}
\usepackage[latin2]{inputenc}
\usepackage{dcolumn,nicefrac,blindtext,nicefrac,pgfplots}

\newcommand{\median}{Q_{\nicefrac 1 2}}
\newcommand{\dQr}{\Delta Q_r}
\newcommand{\Erwt}{\langle t \rangle}
\newcommand{\logdec}{\log_{10}}

\begin{document}
\title{Transport-Generated Signals Uncover Geometric Features of Evolving 
Branched Structures}
\author{Fabian H.\ Kreten}
\affiliation{Department of Theoretical Physics and Center for Biophysics, 
Saarland University, 66123 Saarbr\"ucken, Germany}
\author{Ludger Santen}
\affiliation{Department of Theoretical Physics and Center for Biophysics, 
Saarland University, 66123 Saarbr\"ucken, Germany}
\author{Reza Shaebani}
\email{shaebani@lusi.uni-sb.de}
\affiliation{Department of Theoretical Physics and Center for Biophysics, 
Saarland University, 66123 Saarbr\"ucken, Germany}

\begin{abstract}
Branched structures that evolve over time critically determine the function 
of various natural and engineered systems, including growing vasculature, 
neural arborization, pulmonary networks such as lungs, river basins, power 
distribution networks, and synthetic flow media. Inferring the underlying 
geometric properties of such systems and monitoring their structural and 
morphological evolution is therefore essential. However, this remains a 
major challenge due to limited access and the transient nature of the 
internal states. Here, we present a general framework for recovering the 
geometric features of evolving branched structures by analyzing the signals 
generated by tracer particles during transport. As tracers traverse the 
structure, they emit detectable pulses upon reaching a fixed observation 
point. We show that the statistical properties of this signal intensity--- 
which reflect underlying first-passage dynamics--- encode key structural 
features such as network extent, localized trapping frequency, and bias 
of motion (e.g., due to branch tapering). Crucially, this method enables 
inference from externally observable quantities, requiring no knowledge 
of individual particle trajectories or internal measurements. Our approach 
provides a scalable, non-invasive strategy for probing dynamic complex 
geometries across a wide range of systems.
\end{abstract}

\maketitle
 
Dynamic branched structures play a key role in shaping the function of diverse 
natural and engineered systems by controlling the transport of matter, energy, 
and information. Examples in nature include the nutrient-distributing roots 
and branched bodies of plants, vascular networks, branched organoids, neuronal 
dendrites, the bronchial architecture of lungs, and the drainage patterns of 
river basins \cite{Ball09,Kocillari21,Felici04,Randriamanantsoa22,Spruston08,
Liao21,Jose18,Perron12,DesaiChowdhry22,Grigoriev22,Rinaldo14,Borse23}. 
Technological and synthetic analogs include dendrimer macromolecules, synthetic 
polymer configurations, bioinspired hierarchical porous materials such as 
graphene aerogels, and large-scale infrastructures such as communication 
grids and power distribution networks \cite{Wu12,Zhou24,Zheng17,Banavar99,
Schafer22,Yang22}. Branched architectures also play a crucial role in abstract 
frameworks such as network and graph theory, epidemic modeling, and computational 
search algorithms \cite{PastorSatorras15,Newman10,Kleinberg00,Bollobas04}. 
Despite the central role of internal geometry in determining the function, 
monitoring the structural and morphological evolution of these systems remains 
a major challenge due to limited direct access to their often-hidden structures 
and the transient nature of their adapting geometries. Predicting such structural 
evolution is especially critical in biological contexts--- for instance, to track 
the degeneration of neuronal dendrites in the progression of neurodegenerative 
diseases \cite{Spruston08,Forrest18}.

\begin{figure}[b]
\centering
\includegraphics[width=0.42\textwidth, trim=0 170 10 115, clip]{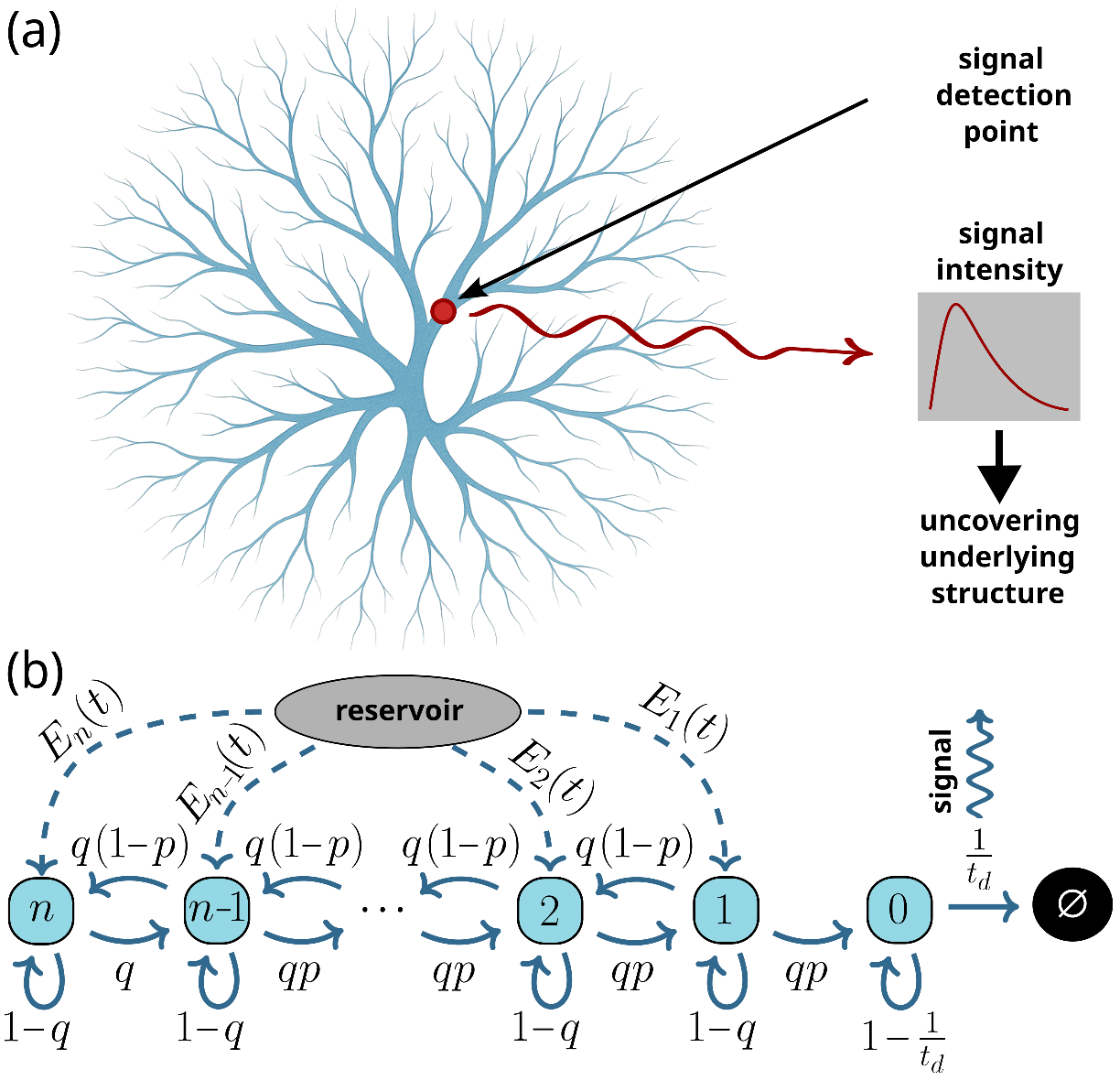}
\caption{(a) Schematic illustration of a branched structure with a 
fixed detection point that registers tracer-generated signals used for structural 
inference. (b) Sketch of the transitions in our model leading to signal generation.}
\label{Fig:1}
\end{figure}

Tracking the diffusive dynamics of tracer particles has long served as a method to 
probe complex, labyrinthine structures by linking stochastic transport statistics 
to underlying geometry \cite{Felici04,benAvraham00,Mair99,Tierno16,Krishna09,Shaebani18,
Cooper16,Agliari07,Sadjadi08}. However, these approaches typically require direct 
observation of individual tracer trajectories, which poses substantial technical 
challenges--- particularly in biological systems--- due to the invasive nature of 
the measurements, the need for high-resolution tracking over complete trajectories, 
and the difficulty of acquiring sufficient statistics within the timescales of 
structural evolution. As an alternative, we propose that tracers need to only emit 
detectable signals upon passing a designated observation point along the structure. 
This circumvents the need for invasive measurements or trajectory reconstruction. 
The resulting aggregate signal--- quantified by its intensity over time--- retains 
key statistical signatures shaped by the underlying geometry. By repeating this 
measurement over time, the structural evolution can be reconstructed. This opens 
a path toward inferring global structural characteristics from localized, 
time-resolved signal measurements.

Here, we develop a general predictive framework to recover the hidden geometric 
features of a broad class of evolving branched structures from the signal intensity 
distribution generated by tracers at a fixed (arbitrarily chosen) detection point 
(see Fig.\,\ref{Fig:1}). By analyzing time-dependent signal statistics (linked to 
first-passage transport dynamics), we identify key signatures of global structural 
properties, including network extent, bias of motion (e.g., due to tapering toward 
or away from the leaves), and localized trapping frequency (arising from boundary 
stickiness, internal traps, or transient caging at junctions and branch segments). 
Importantly, our approach does not require tracking individual particles or continuously 
monitoring all internal nodes of the network, making it well-suited for minimally 
invasive probing of complex structures. The framework applies broadly across physical, 
biological, and synthetic branched systems where stochastic transport processes 
and signal emissions are present. We also discuss the feasibility of signal-based 
detection at biologically relevant scales, such as in neuronal dendrites, to 
demonstrate the experimental plausibility of our approach.

\emph{Branched Structure Model.---} We adopt a coarse-grained perspective and consider 
a branched tree structure consisting of $n$ generations of junctions and a linear 
extent $n L$, where $L$ is the average distance between adjacent nodes. Noninteracting 
tracer particles perform stochastic hopping between the nodes of the tree \cite{Jose18}. 
Initially, a reservoir contains $\mathcal{N}$ tracer particles. Each tracer enters 
the tree at a random node with depth index $i\,{\in}\,[0,n]$, where $i\,{=}\,0$ 
corresponds to the main root and $i\,{=}\,n$ to the leaves. Entry occurs at a 
random time drawn from a geometric distribution with mean $t_e$, and the entry 
position is uniformly distributed over the network nodes, leading to an exponentially 
growing entry probability with the node depth. This leads to a joint distribution 
for the entry depth and time, given by $E_{i}(t)\,{=}\,\frac{1}{t_e}\!\left(1 {-} 
\frac{1}{t_e}\right)^t\!\!\frac{2^{i{-}1}}{2^n{-}1}$ in a binary tree. This injection 
process is illustrated by dashed arrows in Fig.\,\ref{Fig:1}(b).

\begin{figure*}[t]
\includegraphics[width=0.99\textwidth, trim=0 0 0 0, clip]{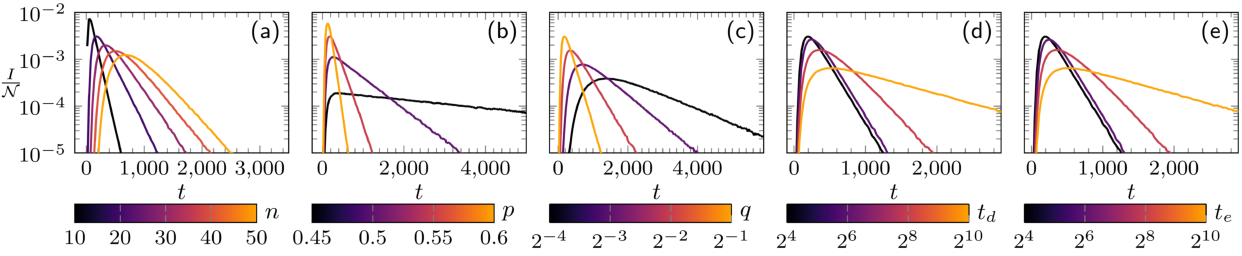}
\caption{Time evolution of the overall signal intensity $I(t)$, scaled by the reservoir 
capacity $\mathcal{N}$ and averaged over bins of 30 time steps width, for different 
values of the model parameters. The reference parameter values are taken to be $n\,
{=}\,20$, $p\,{=}\,0.55$, $q\,{=}\,0.5$, and $t_e\,{=}\,t_d\,{=}\,1$. The overall 
signal intensity is compared for different values of (a) $n$, (b) $p$, (c) $q$, (d) 
$t_d$, and (e) $t_e$. The selected values for each parameter are given in the color 
box of the respective panel.}
\label{Fig:2}
\end{figure*}

We consider dynamics resolved at a coarse-grained time scale $\Delta t$, representing 
the resolution of observation or measurement. At each time increment $\Delta t$, a 
tracer either hops randomly to one of its neighboring nodes with probability $q$, 
or remains at its current node with probability $1{-}q$. This spontaneous hopping 
corresponds to a geometrically distributed waiting time at each node, which captures 
both transient stochastic caging along the tree structure and localized trapping 
events caused by sticky or roughened boundaries. Extending the framework to systems 
with heavy-tailed waiting time distributions is possible, allowing the modeling of 
anomalous or non-Markovian transport. The hopping probability $q$ can be related to 
the underlying physical properties through the relation $\Delta t{/}q\,{=}\,\Erwt$, 
where $\Erwt$ is the mean escape time from a node (e.g., a junction) to any 
neighboring furcation in real time. Physically, $\Erwt$ is proportional to 
$\omega_m{/}(\omega_m{+}\omega_w)$ in systems with stochastic binding to sticky 
boundaries, where $\omega_m$ and $\omega_w$ denote the unbinding and binding 
rates, respectively. In geometrically caged environments, $\Erwt$ instead scales 
with $1{+}\frac{V_\text{cages}}{V_\text{channel}}$, where $V_\text{cages}$ is the 
total volume of localized traps and $V_\text{channel}$ is the volume of accessible 
transport paths; see \emph{Suppl.\,Info.}\ and \cite{Dagdug07}. This mapping ensures 
that the mean local waiting time in the coarse-grained model is matched to that 
in the underlying physical process.

A topological bias parameter $p$ is introduced to account for directional preference in 
tracer motion. At each step, a tracer hops toward the root with probability $p$, or toward 
the leaves with probability $1\,{-}\,p$. This bias can arise from an imposed flow through 
the structure, or in the case of purely diffusive dynamics, from geometrical asymmetries 
such as tapering of the channel cross-section. Moreover, for diffusive motion on a symmetric 
tree composed of cylindrical branches, each furcating into $\mathfrak z\,{-}\,1$ children, 
and neglecting complications at the branch points, the directional bias can be approximated 
by $p\,{=}\,\frac{A_p}{A_p\,{+}\,(\mathfrak z\,{-}\,1)A_c}$, where $A_p$ and $A_c$ denote 
the cross-sectional areas of the parent and child branches, respectively. The areas are 
linked via the allometric relation $d_p^{\,\kappa}\,{=}\,\sum_{i{=}1}^{\mathfrak z\,{-}\,1} 
d_{c}^{\,\kappa}$, where $d_p$ and $d_{c}$ are the diameters of the parent and child branches 
at a furcation, and $\kappa$ is an allometric exponent. Empirical studies suggest representative 
values of the allometric exponent, with $\kappa\,{=}\,\frac 3 2$ for neuronal dendrites, 
$\kappa\,{=}\,2$ for botanical trees, and $\kappa\,{=}\,3$ for vascular and pulmonary networks 
\cite{Liao21,Grigoriev22,Murray26,Rall59}. Assuming bifurcations ($\mathfrak z\,{=}\,3$) 
and symmetric daughter branches yields estimated values of $p\,{\simeq}\,0.56$, $0.5$ and 
$0.44$, respectively, in these cases. For general channel profiles and driving forces, $p$ 
can be calculated by solving a Fick-Jacobs-like equation; see \emph{Suppl.\,Info.}\ for details.

As a proof of concept, we focus here on regular tree geometries, i.e.\ with uniform model 
parameters ${n,q,p}$. However, the conclusions of our framework are expected to hold under 
moderate (realistic) levels of structural irregularity, including local fluctuations in 
connectivity and global variations in branching statistics. This robustness stems from 
the fact that the signal characteristics we analyze are ultimately governed by first-passage 
dynamics--- quantities we have previously shown to be resilient against such imperfections, 
e.g.\ in the context of transport on neuronal dendritic trees \cite{Jose18,Shaebani18}.

\emph{Signal Generation.---} After a tracer reaches the designated detection point, it emits 
a transient pulse following a random delay. This delay, referred to as the \emph{emission time}, 
is drawn from a geometric distribution with mean $t_d$. The inclusion of this delay accounts 
for possible local activation processes--- such as biochemical or electrical mechanisms--- 
that may be required to generate a detectable signal upon arrival. These pulses accumulate 
into the total observable signal $I(t)$, from which we infer the model parameters $(n,p,q)$. 
$I(t)$ thus reflects the combined statistics of three stochastic processes: the entry time, 
the first-passage time through the network, and the emission delay. In the special case 
$t_e\,{=}\,t_d\,{=}\,1$, $I(t)$ reduces--- up to a scaling factor and time shift--- to 
the first-passage time distribution. We provide an analytical connection between the model 
parameters $(n,p,q)$ and first-passage quantities elsewhere (see also \cite{Jose18,Shaebani18}); 
however, these quantities are not directly accessible in practice, as they require tracking 
individual tracers. Instead, the present work develops an inference framework grounded in 
processing the generated signal $I(t)$, which is externally measurable.

The tracer injection protocol and emission statistics used here are illustrative and can 
be adapted to suit system-specific constraints. While the location of the detection point 
is, in principle, arbitrary, more central placements generally yield more informative 
signals and are analytically more tractable. The underlying geometry and boundary conditions 
may also vary depending on the application. Here, by selecting the root of the tree as the 
detection point, we map the transport problem onto an effective one-dimensional lattice--- 
a reduction commonly employed in the analysis of finite Cayley trees and infinite Bethe 
lattices \cite{Agliari08,Metzler14,Shaebani18,Redner01,Hughes82,Skarpalezos13}. In this 
setting, the signal is given by $I(t{+}1) \,{=}\,P_0(t){/}t_d$, where $P_0(t)$ 
is the occupation probability of the root at time $t$. This relation directly links the 
temporal signal to the transport dynamics through the network. Figure~\ref{Fig:1}(b) 
illustrates the corresponding effective one-dimensional lattice representation. Solid 
arrows indicate transition probabilities between neighboring sites, while dashed arrows 
represent the total probability influx from external sources. The full dynamics are 
described by a set of coupled master equations, provided in the \emph{Suppl.\,Info.}

\begin{figure*}[t]
\includegraphics[width=0.99\textwidth, trim=0 0 0 0, clip]{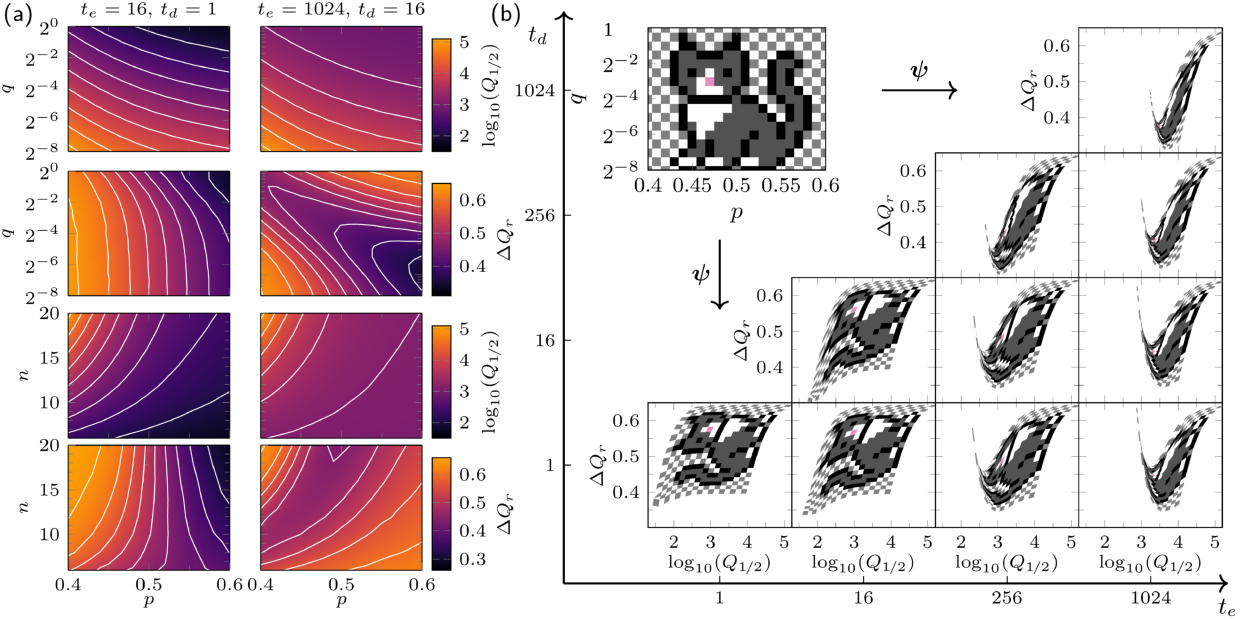}
\caption{(a) Heatmaps of $\log_{10}(\median)$ and $\dQr$ in the $(p,q)$-section (upper 
panels, $n=10$) and $(p,n)$-section (lower panels, $q=0.5$) of the model parameter space, 
shown for low (left column) and high (right column) values of $t_e$ and $t_d$. (b) Effect 
of $t_e$ and $t_d$ on the mapping $\boldsymbol\psi$ from model parameters $(p,q)$ to shape 
parameters $(\log_{10}(\median), \dQr)$, with $n=10$. The local neighborhood of each sampled 
model parameter set is color-coded (inset), and corresponding mapped regions in shape parameter 
space are shown in the small panels. Due to symmetry under swapping $t_e\,{\leftrightarrow}
\,t_d$, only cases with $t_e\,{\geq}\,t_d$ are shown.}
\label{Fig:3}
\end{figure*}

\emph{Signal Processing.---}  To assess whether the signal $I(t)$ carries sufficient information 
to infer the model parameters, we performed extensive Monte Carlo simulations of the outlined 
dynamics, starting from a reservoir of $\mathcal N\,{=}\,10^6$ tracer particles. Figure \ref{Fig:2} 
shows a typical temporal evolution of $I(t)$, which exhibits a prominent peak followed by an 
exponential decay at long times. The location and height of the peak, as well as the decay 
rate of the tail, depend systematically on the model parameters. Broadly, the signal becomes 
wider and decays more slowly--- with a lower peak and a delayed maximum--- for increasing tree 
depth $n$, decreasing bias $p$ or hopping probability $q$, or increasing injection/emission 
delays $t_e$ and $t_d$ (see Fig.\,\ref{Fig:2}). Notably, $I(t)$ is invariant under the exchange 
of $t_e$ and $t_d$; the longer of the two timescales dominates the asymptotic behavior of $I(t)$.

To quantify these dependencies and evaluate which parameters can be extracted from signal 
observations, we computed a set of shape parameters of $I(t)$. When $t_e$ and $t_d$ are moderate--- 
that is, when the entry and emission timescales are sufficiently short compared to the mean 
first-passage time required for tracers to explore the structure--- these parameters fall 
into two distinct behavioral classes in the $(n,p,q)$-space. For large $t_e$ and $t_d$, however, 
this distinction breaks down, and parameter identifiability becomes limited. Figure \ref{Fig:3}(a) 
presents the logarithm of the signal median, $\logdec(\median)$, across $(p,q)$ and $(p,n)$ 
sections of the parameter space for both low and high values of $t_e$; see also Suppl.\,Fig.\,S2 
for other $t_e$ and $t_d$ values. This observable varies smoothly and monotonically with the 
model parameters. Increasing $t_e$ and $t_d$ shifts the median to later times without altering 
its overall landscape. Similar trends are observed for the signal mean, higher moments, and 
peak height (not shown). A representative of the second class is the normalized interquartile 
range, $\dQr\,{=}\,\frac{Q_{\nicefrac 3 4}\,{-}\,Q_{\nicefrac 1 4}}{Q_{\nicefrac 3 4}\,{+}\,
Q_{\nicefrac 1 4}}$, where $Q_{\nicefrac i 4}$ denotes the $i$-th quartile. This parameter 
measures the dispersion of $I(t)$. At low $t_e$ and $t_d$, $\dQr$ behaves smoothly and 
monotonically, but for larger time scales, nonmonotonic patterns and regions of near-degeneracy 
emerge, where $\logdec(\median)$ and $\dQr$ become nearly colinear. Other observables in this 
class include the normalized variance, skewness, and kurtosis of $I(t)$. The identification 
of only two effective classes of shape parameters implies that these signal measurements 
alone constrain the model to a one-dimensional manifold in the $(n,p,q)$ parameter space. 
Knowledge of one parameter (or a constitutive relation among them) is then sufficient to 
infer the others uniquely.

To examine the roles of $t_e$ and $t_d$, and to identify regions of (local) invertibility of the 
mapping $\boldsymbol \psi$ from model to shape parameters, we simulated $I(t)$ across a regular 
grid in parameter space and extracted $\logdec(\median)$ and $\dQr$. Figure \ref{Fig:3}(b) 
visualizes this mapping for a representative $(p,q)$ section. At small $t_e$ and $t_d$, the 
map is nearly one-to-one and preserves structure under smooth deformations (e.g., flips and 
rotations). As $t_e$ and $t_d$ increase, regions of the model space--- particularly at high 
$q$--- collapse into narrow bands in the shape parameter domain, signaling the breakdown of 
parameter extractability. Similar behavior is observed in other sections through $(n,p,q)$-space; 
see e.g.\ Suppl.\,Fig.\,S3.

\emph{Invertibility and Experimental Bounds.---} As shown above, $I(t)$ encodes structural 
information about the tree, and we have verified that, under moderate entering and emission 
timescales, two of the three core model parameters $(n,p,q)$ can be inferred from shape 
parameters of the signal. To uniquely recover all three parameters, an additional 
independent observable is required. This could come from higher-order shape parameters. 
Alternatively, statistical approaches such as maximum likelihood estimation, Kullback-Leibler 
divergence minimization, or empirical characteristic function analysis can be used. The 
injection and emission timescales $t_e$ and $t_d$ could also be included as estimable 
parameters. This framework provides a foundation for noninvasive structural probing of 
branched systems, including delicate biological structures such as neuronal dendrites. 

A key limitation, however, is the breakdown of invertibility in the high-$q$ regime when 
$t_e$ or $t_d$ become large. This reflects a fundamental constraint: the timescales of 
signal generation must be compatible with those of the underlying transport dynamics. 
To quantify this, we identify a critical threshold $q^*$ beyond which signal-based 
inference fails. Based on simulated data and a minimum-distance threshold in the 
$(\logdec(\median),\dQr)$-domain, we find $q^* {\simeq} \sqrt{\nicefrac{\Delta t}{
\max(t_e, t_d)}}$. To fully resolve the model across the entire $(n,p,q)$-space, 
this threshold must exceed the upper bound for $q$, given by $q \,{=}\, \Delta t
\frac{2D}{L^2}$, corresponding to trap-free diffusion. This sets an upper bound 
\begin{equation}
\nonumber
\Delta t \,{\leq}\, \frac{L^4}{4D^2\max(t_e, t_d)},
\end{equation}
on the temporal resolution required to maintain invertibility. This condition is 
consistent with the restrictions that the time resolution must satisfy 
$\Delta t \,{\leq}\,\frac{L^2}{2D}$ for $q \,{\leq}\, 1$, and the mean 
time $\frac{2D}{L^2}$ to diffuse between adjacent nodes must exceed $\max(t_e, t_d)$.

We have presented a general framework for inferring the geometric features of 
evolving branched structures from externally measurable, transport-generated 
signals. We showed how key structural properties--- such as network extent, 
directional bias, and localized trapping--- can be extracted without tracking 
individual tracers. This scalable and minimally invasive approach applies 
broadly across biological, physical, and synthetic systems. The concept of 
signal emission upon reaching a detection site is not merely theoretical. For 
instance, we demonstrate elsewhere that brain-targeted drug delivery can deploy 
multilamellar liposomes into neuronal dendrites, where they traverse complex 
arborizations, reach the soma, and release mRNA cargo. Ribosomes translate 
this mRNA into proteins whose concentration--- summed over many neurons--- 
is externally measurable by the MRI technique, effectively realizing a signal 
analogous to $I(t)$ at biologically relevant scales. Our approach thus also 
offers a diagnostic tool for probing the architecture of evolving biological 
media. While our study focused on tree-like networks, the core ideas are 
extendable to more complex architectures, including graphs with loops, 
irregular or weighted topologies, and hybrid tree-graph structures. Future 
directions include accounting for noisy or incomplete data, incorporating 
interacting tracers and feedback-coupled dynamics, and integrating statistical 
inference or machine learning methods to broaden the applicability and 
robustness of this framework.

This work was supported by the Deutsche Forschungsgemeinschaft (DFG) within the 
collaborative research center SFB 1027 and also via grants INST 256/539-1, which 
funded the computing resources at Saarland University. R.S.\ acknowledges support 
by the Young Investigator Grant of Saarland University, Grant No.\ 7410110401.

\bibliography{Refs-Tree}

\newpage

\begin{widetext}

\vspace{20mm}

\begin{center}
{\LARGE Supplementary Information to Transport-Generated Signals Uncover Geometric 
Features of Evolving Branched Structures}\\
\end{center}

\section{Mapping Model Parameters $p$ and $q$ to Channel Geometry}

To map the bias probability $p$ and hopping probability $q$ in our coarse-grained 
model to the underlying channel geometry, we analyze the dynamics of a tracer 
particle confined to the channel segment between neighboring parent and child 
intersections. To maintain tractability, we first derive an effective 1D 
Fick-Jacobs equation using the approach of Zwanzig \cite{Zwanzig92} in 
Subsec.\,A. The parameters $p$ and $q$ correspond to the splitting probabilities 
and mean escape times, respectively, which can be obtained using standard techniques, 
since the effective 1D description is a Smoluchowski equation \cite{Gardiner85,
vanKampen11}. In Subsec.\,B, we compute these quantities for general channel 
profiles and under additional driving forces. Subsection\;C illustrates the results 
by evaluating $p$ and $q$ for trees composed of cylindrical segments with diameter 
changes at the furcations, both in the absence of forces and under constant drift. 
Subsection\;D addresses how the presence of traps in form of cavities modifies the 
effective hopping probability $q$.

\subsection{A.\ Derivation of the effective 1D equation}
\label{sec:derivation}

We start from the full-dimensional Smoluchowski equation
\begin{equation}
\label{eq:smoluchowski}
\frac{\partial}{\partial t} C(t, \bm r) = D \, \nabla{\cdot}\Big(\text{e}^{-\beta 
U(\bm r)} \nabla \! \big(\text{e}^{\beta U(\bm r)}C(t, \bm r)\big)\Big),
\end{equation}
governing the probability $C$ for a tracer particle to be at point $\bm r$ at time $t$.
The potential is composed of two terms, $U(\bm r) = U_0(\bm r) +V(\bm r)$, with
\[ U_0 = \begin{cases} 0 &,\ \bm r \in \text{channel} \\ \infty &,\ \text{else}\end{cases} \]
restricting the residence of the particle to the channel volume. $V(\bm r)$ is the potential 
of additional forces acting on the particle--- which should be constant when restricted 
to any channel segment and its values in the child channel should be identical for 
identical distances from the furcation.
\smallskip\smallskip\smallskip

We denote the distance of the channel cross-section from the intersection (measured 
along the channel axis) with $\xi$. Thus, the parent intersection is at $\xi\,{=}\,0$ 
and the child intersections are at $\xi\,{=}\,{-L}$. Then, the marginal probability \[ c(t, \xi) 
= \int_{\mathcal A(\xi)}\!\!C(t, \bm r) \, \text{d}^2\bm r \]
can be defined, where $\mathcal A(\xi)$ is a surface that perpendicularly intersects 
\emph{all} channels at distance $\xi$. The branched channel, the mapping to $\xi$, 
and examples of $\mathcal A(\xi)$ are shown in Fig.\,S1.
\smallskip\smallskip\smallskip

\begin{figure}
\begin{center}
\includegraphics[width=0.4\textwidth]{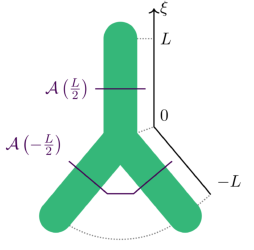} \\
\end{center}
Suppl.\,Fig.\,S1: Sketch of one branching point of the channel and its mapping to 
the 1D interval $\xi\,{\in}\,[-L,L]$. The space occupied by the channels is shown 
in green. The dotted lines indicate the branching points along the channel with 
their corresponding $\xi$ values. Two surfaces $\mathcal A$ perpendicularly 
intersecting the channels, one above and one below the furcation, are indicated 
by purple solid lines.
\end{figure}

Applying $\int_{\mathcal A}\text{d}^2{\bm r}$ on both sides of Eq.\,(\ref{eq:smoluchowski}), 
expressing the gradient and divergence in local coordinates along and perpendicular 
to the channel axis, and using the Gaussian theorem with $\text{e}^{-\beta U}{\equiv}\,0$ 
outside and on the channel walls yields
\begin{equation}
\label{eq:smoluchowski_1st_app}
\frac{\partial}{\partial t} c = D\frac{\partial}{\partial \xi}\int_{\mathcal A(\xi)}\!\!
\text{e}^{-\beta U}\frac{\partial}{\partial \xi}\Big(\text{e}^{\beta U}C(t,\bm r)\Big)
\text{d}^2\bm r.
\end{equation}
\smallskip\smallskip\smallskip

We further assume local equilibrium on each segment through all channels, which is 
justified for long channels. Introducing the local partition function $\mathcal Z(\xi)
\,{=}\,\int_{\mathcal A}\text{e}^{-\beta U(\bm r)}\text{d}^2\bm r\,{=}\,A(\xi)\,
\text{e}^{-\beta V(\xi)}$ allows the approximation $C(t, \bm r)\,{\simeq}\,c(t, \xi)
\frac{\text{e}^{-\beta U(\bm r)}}{\mathcal Z(\xi)}$. By substituting it in 
Eq.\,(\ref{eq:smoluchowski_1st_app}), it can be shown that $c(t,\xi)$ obeys the 
one-dimensional Smoluchowski equation
\begin{equation}
\label{eq:smoluchowski_2nd_app}
\frac{\partial}{\partial t} c = D\frac{\partial}{\partial \xi}\Big(\mathcal Z(\xi)
\frac{\partial}{\partial \xi}\frac{c}{\mathcal Z(\xi)}\Big) = D\frac{\partial^2}{
\partial \xi^2}c-D\frac{\partial}{\partial \xi}\Big(\frac{\mathcal Z'}{\mathcal Z}c\Big).
\end{equation}
In the absence of additional forces, the local partition function is given by the 
total channel cross section $A(\xi)$ at distance $\xi$, $\mathcal Z(\xi)\,{=}\,A(\xi)$, 
and Eq.\,(\ref{eq:smoluchowski_2nd_app}) reduces to the Fick-Jacobs equation
\begin{equation}
\label{eq:fick_jacobs_eq}
\frac{\partial}{\partial t} c = D\frac{\partial}{\partial \xi}\Big( A(\xi)\frac{
\partial}{\partial \xi}\frac{c}{A(\xi)}\Big)= D\frac{\partial^2}{\partial \xi^2}c
-D\frac{\partial}{\partial \xi}\Big(\frac{A'}{A}c\Big).
\end{equation}

\subsection{B.\ Calculation of splitting probabilities and mean escape time}
\label{sec:splitting_escape}

Since Eq.\,(\ref{eq:smoluchowski_2nd_app}) is a Smoluchowski equation, the splitting 
probabilities and mean escape time can be derived generally. To facilitate this, we 
shift the coordinate system such that the origin lies at the end of the child channels, 
placing the intersection at $\xi\,{=}\,L$ and the end of the parent channel at $\xi\,{=}
\,{2L}$. To reflect this change, we denote the spatial coordinate by $x$, replacing the 
earlier notation $\xi$. For convenience, we introduce dimensionless coordinates $\tilde x
\,{=}\,\frac x L$ and $\tilde t\,{=}\,t\frac{D}{L^2}$; however, we explicitly mark them 
with tildes only when they appear alongside their dimensional counterparts.
\smallskip\smallskip\smallskip

We define the transport operator $\ensuremath\mathsf{L}$ as $\ensuremath\mathsf{L} c\,{=}\,
\frac{\partial}{\partial x}\Big(\mathcal Z(x)\frac{\partial}{\partial x}\frac{c(x)}{\mathcal 
Z(x)}\Big)$. Then, the splitting probabilities $w_\pm(x)$--- that the parent (child) 
intersection is reached before the child (parent) intersection, starting from $x$--- 
follow $\ensuremath\mathsf{L}^\dagger w_\pm = 0$, where $\ensuremath\mathsf{L}^\dagger$ 
is the adjoint of $\ensuremath\mathsf{L}$. This results in the ordinary differential equation
\begin{equation}
\label{eq:deq_splitting_prob}
\frac{1}{\mathcal Z}\frac{\text{d}}{\text{d}x}\Big(\mathcal Z\frac{\text{d}}{\text{d}x} 
w_\pm\Big) = 0.
\end{equation}
Obviously, $w_+ + w_- = 1$ and $w_\pm$ has to fulfill $0 = w_+(0) = w_-(2)$ and $1 = w_-(0) = 
w_+(2)$. The solutions of Eq.\,(\ref{eq:deq_splitting_prob}) respecting these boundary conditions are
\begin{equation}
\label{eq:sol_splitting_probs}
w_+(x) = \frac{\int_{0}^{x}\mathcal Z^{-1}(x')\text{d}x'}{\int_{0}^{2}\mathcal Z^{-1}(x')\text{d}x'}
\quad\text{and}\quad
w_-(x) = \frac{\int_{x}^{2}\mathcal Z^{-1}(x')\text{d}x'}{\int_{0}^{2}\mathcal Z^{-1}(x')\text{d}x'}.
\end{equation}
The bias probability $p$ is the probability that a particle starting at the intersection reaches 
the parent intersection before any child intersection, hence
\begin{equation}
\label{eq:calibration_p}
p = w_+(1) = \frac{\int_{0}^{1}\mathcal Z^{-1}(x')\text{d}x'}{\int_{0}^{2}\mathcal Z^{-1}(x')\text{d}x'}.
\end{equation}
Depending on the specific geometry, the functional form of $\mathcal Z$ varies with the level of the 
intersection, resulting in a depth-dependent bias parameter $p$. The model described by Master 
Eqs.\,(\ref{Eq:MasterEqs}) can be readily extended to incorporate such depth-dependent bias. 
Alternatively, a single effective value of $p$ can be used.
\smallskip\smallskip\smallskip

The mean escape time $t(x)$ starting from $x$ has to fulfil the differential equation
\begin{equation}
\label{eq:deq_escape_time}
\ensuremath\mathsf{L}^\dagger t =
\frac{1}{\mathcal Z}\frac{\partial}{\partial x}\Big(\mathcal Z\frac{\partial}{\partial x}
t(x)\Big) = -1,
\end{equation}
and the boundary conditions $t(0) \,{=}\, 0 \,{=}\, t(2)$. Since $w_\pm$ are independent 
solutions of Eq.\,(\ref{eq:deq_splitting_prob}), the homogeneous part of the solution of 
Eq.\,(\ref{eq:deq_escape_time}) is of the form $t(x) = B_+(x)w_+(x) +B_-(x)w_-(x)$. After 
some algebra one finds that
\begin{equation}
\label{eq:sol_escape_time}
t(x) = 
w_+(x) \int_{x}^{2}\frac{\int_{0}^{x'}\mathcal Z(x'')\,\text{d}x''}{\mathcal Z(x')}\text{d}x' 
-w_-(x) \int_{0}^{x}\frac{\int_{0}^{x'}\mathcal Z(x'')\,\text{d}x''}{\mathcal Z(x')}\text{d}x'
\end{equation}
is the solution of Eq.\,(\ref{eq:deq_escape_time}) respecting the boundary conditions 
$t(0) \,{=}\, 0 \,{=}\, t(2)$ \cite{Gardiner85}. Equation\;(\ref{eq:sol_escape_time}) 
provides the dimensionless escape time $\tilde t(\tilde x)$. The corresponding dimensional 
escape time and the moving probability are then given by
\begin{equation}
\label{eq:sol_escape_time_dimfull}
t(x) = \frac{L^2}{D}\tilde t\left(\frac x L\right)
\end{equation}
and 
\begin{equation}
\label{eq:moving_prob}
q = \Delta t\frac{1}{t(L)}
\end{equation}
where $\Delta t$ is the coarse-grained time resolution. This ensures that the mean waiting 
times at intersections match in both the physical (real-time) and the coarse-grained model 
description.

\subsection{C.\ Examples}
\label{sec:examples}

Next, we calculate the bias and hopping probabilities $p$ and $q$ for two representative 
examples. In the first case, we apply the general formulas derived in the previous section. 
In the second example, we explicitly solve the differential equations governing the splitting 
probabilities and mean escape time from first principles.

\subsubsection{1.\ Constrictions at furcations}
\label{sec:example_constricting_furcations}

The first example is a tracer particle moving through a tree composed of cylindrical channels, 
in the absence of external forces. At each furcation, a parent branch with cross-sectional 
area $A_p$ splits into $\mathfrak z\,{-}\,1$ child branches, each with cross-sectional area 
$A_c$. Notably, the parent area $A_p$ at any given branching point corresponds to the child 
area $A_c$ of the preceding furcation. The analysis that follows is valid for both widening 
($A_c \leq A_p$) and tapering ($A_c \geq A_p$) geometries.
\smallskip\smallskip\smallskip

Expressed in terms of the dimensionless position, the local partition function reads
\[ \mathcal Z(x) = \begin{cases} (\mathfrak z -1)A_c &,\ 0 \leq x < 1 \\ A_p &,\ 1 \leq x 
\leq 2 \end{cases}. \]
With the integrals $\int_{0}^{1}\frac{1}{\mathcal Z(x)}\text{d}x = \frac 1 {(\mathfrak z -1)A_c}$ 
and $\int_{1}^{2}\frac{1}{\mathcal Z(x)}\text{d}x = \frac 1 {A_p}$, Eq.\,(\ref{eq:calibration_p}) 
can be evaluated. After some algebra one finds
\[
\begin{aligned}
w_+(1) &= \frac{A_p}{A_p +(\mathfrak z -1)A_c} = p, \\
w_-(1) &= \frac{(\mathfrak z -1)A_c}{A_p +(\mathfrak z -1)A_c}.
\end{aligned}
\]
The prefactors in Eq.\,(\ref{eq:sol_escape_time}) for $x\,{=}\,1$ read
\[
\int_{1}^{2}\frac{\int_{0}^{x}\mathcal Z(x')\text{d}x'}{\mathcal Z(x)}\text{d}x = 
\frac{(\mathfrak z -1)A_c}{A_p} +\frac 1 2
\quad\text{and}\quad
\int_{0}^{1}\frac{\int_{0}^{x}\mathcal Z(x')\text{d}x'}{\mathcal Z(x)}\text{d}x = 
\frac 1 2,
\] 
thus, the dimensionless mean escape time is $t(1) = \frac 1 2$. Redimensionalizing 
yields $t(L) = \frac{L^2}{2D}$, identical to the escape time for a straight tube 
of length $2L$, independent of branching and tapering. The hopping probability is
\begin{equation}
\label{eq:moving_prob_constricting_furcations}
q = \Delta t\frac{2D}{L^2}.
\end{equation}

\subsubsection{2.\ Constant drift}
\label{sec:example_constant_drift}

As a second example, we consider a diffusing particle subject to a constant drift velocity 
$v$. Constant drift due to hydrodynamic flow requires that the cross-sectional area be 
preserved before and after the intersection due to mass conservation. However, the results 
derived in this subsection apply to all combined effects of geometry and driving forces 
that yield a local partition function satisfying $\frac{\mathcal Z'}{\mathcal Z} = v$. 
The constant-drift case can also serve as an approximation for more complex scenarios, 
provided that $\frac{\mathcal Z'}{\mathcal Z}$ is constant in lowest order. In such 
cases, the results apply only up to the first order in the drift velocity $v$.
\smallskip\smallskip\smallskip

In contrast to the previous example, we solve Eq.\,(\ref{eq:deq_splitting_prob}) and 
Eq.\,(\ref{eq:deq_escape_time}) directly rather than plugging $\mathcal Z = \text{e}^{vx}$ 
into Eq.\,(\ref{eq:calibration_p}) and Eq.\,(\ref{eq:moving_prob}). As before, we use 
dimensionless coordinates, i.e., measure all lengths in units of $L$ and all times 
in units of $\frac{L^2}{D}$. Consequently, the velocity $v$ is measured in units of 
$\frac D L$ and the dimensionless velocity is the Peclet number.
\smallskip\smallskip\smallskip

In the present case, the splitting probabilities $w_\pm$ follow the ordinary differential 
equation
\begin{equation}
\label{eq:deq_splitting_prob_drift}
\frac{\text{d}^2}{\text{d} x^2} w_\pm(x) + v \frac{\text{d}}{\text{d} x} w_\pm(x) = 0,
\end{equation}
with the general solution
\begin{equation}
\label{eq:sol_splitting_prob_drift}
w_\pm(x) = A +B\text{e}^{-vx}.
\end{equation}
The splitting probability to the parent intersection $w_+$ has to fulfil the boundary 
conditions $w_+(0) \,{=}\, 0$ and $w_+(2) \,{=}\, 1$, yielding $w_+(x) \,{=}\, 
\frac{1 -\text{e}^{-vx}}{1 -\text{e}^{-2v}}$. Then, after some algebra, the bias 
probability can be obtained as
\begin{equation}
\label{eq:bias_drift}
p = w_+(1) = \frac 1 2\left(1 +\tanh\frac v 2\right).
\end{equation}
\smallskip\smallskip\smallskip

The mean escape time starting from $x$, $t(x)$, obeys
\begin{equation}
\label{eq:deq_escape_drift}
\frac{\text{d}^2}{\text{d} x^2}t(x) + v \frac{\text{d}}{\text{d}x}t(x) = -1.
\end{equation}
A particular solution of Eq.\,(\ref{eq:deq_escape_drift}) is $t_p(x) = -\frac x v$, and the 
homogeneous part is the same as Eq.\,(\ref{eq:deq_splitting_prob_drift}). The general solution 
is therefore given by
\begin{equation}
t(x) = A +B \text{e}^{-vx} -\frac x v.
\end{equation}
Using the boundary conditions $t(0) = 0 = t(2)$ then yields
\begin{equation}
t(x) = \frac{\text{e}^v \left(1 -\text{e}^{-vx}\right)}{v\sinh v} -\frac x v,
\end{equation}
and the mean escape time starting from the intersection is
\begin{equation}
\label{eq:dedim_escape_drift}
t(1) = \frac{\tanh\frac v 2}{v}
\simeq
\begin{cases}
\frac 1 2 -\frac{v^2}{24} &,\ |v| \ll 1 \\
\frac 1 {|v|}          &,\ |v| \gg 1
\end{cases}.
\end{equation}
\smallskip\smallskip\smallskip

Redimensionalizing Eq.\,(\ref{eq:dedim_escape_drift}) then yields
\begin{equation}
t(1)
= \frac{L}{v} \tanh\left(\frac{vL}{2D}\right)
\simeq
\begin{cases}
\frac{L^2}{2D} -\frac{v^2L^4}{24D^3} &,\ |v| \ll \frac{D}{L} \\
\frac L {|v|}          &,\ |v| \gg \frac{D}{L}
\end{cases}
\end{equation}
and in turn the moving probability reads
\begin{equation}
q = \frac{v\Delta t}{L\tanh\left(\frac{vL}{2D}\right)}
\simeq
\begin{cases}
\frac{24 D^3 \Delta t}{12L^2 D^2 -v^2L^4} &,\ |v| \ll \frac{D}{L} \\
\frac{|v| \Delta t}{L}          &,\ |v| \gg \frac{D}{L}
\end{cases}.
\end{equation}

\subsection{D.\ Incorporating entrapment in cavities}
\label{sec:cavities}

So far, we have assumed a smooth channel in all derivations, i.e., the 
possibility of being trapped is not incorporated into the calibration 
calculations. Note that transient trapping events along the channel do 
not induce any bias in the motion towards one end of the channel segment, 
thus, no modification is required in Eq.\,(\ref{eq:calibration_p}) for the 
parameter $p$. However, for the escape time $t(x)$ given by 
Eq.\,(\ref{eq:sol_escape_time_dimfull}), frequent interruption of motion 
by entrapment events in cavities has a considerable impact. To keep 
the model traceable, this impact is taken into account by an effective 
asymptotic diffusion constant $D_\text{eff}$. Previous studies have 
already calculated such an effective diffusion constant in a geometry 
almost tailored to the diffusive transport in channels studded with 
cavities \cite{Dagdug07}. The cavities were modeled as spheres connected 
to the main channel by narrow cylindrical necks. The effective diffusion 
constant derived in \cite{Dagdug07} is given by
\begin{equation}
\label{eq:effective_D}
D_\text{eff} = D\frac{V_\text{channel}}{V_\text{channel} + V_\text{cages}},
\end{equation}
where $D$ is the diffusion constant without protrusions, $V_\text{channel}$ 
the channel volume, and $V_\text{cages}$ the total volume of cavities.
Strictly, Eq.\,(\ref{eq:effective_D}) is only valid in the absence of driving 
forces causing drift, i.e., only for the first example of diameter change at 
furcations. In presence of drift, like in the second example, corrections to 
the drift velocity are necessary and the effective diffusion coefficient has 
a more complicated functional form \cite{Berezhkovskii11}.
\smallskip\smallskip\smallskip

Substituting $D_{\text{eff}}$ for $D$ in Eq.\,(\ref{eq:moving_prob_constricting_furcations}) 
for $q$ yields
\begin{equation}   
q = \Delta t \frac{2D}{L^2}\frac{V_\text{channel}}{V_\text{channel} +V_\text{cavities}}.
\end{equation}
As $q$ is a probability, it cannot be larger than one, imposing a constraint 
on the time resolution of observation $\Delta t$. Since the relation 
$\frac{V_\text{channel}}{V_\text{channel} +V_\text{cavities}} \leq 1$ always holds, 
the condition
\begin{equation}
\Delta t \leq \frac{L^2}{2D}
\end{equation}
ensures that $q$ always remains as a valid probability, i.e., $q\,{\leq}\,1$.

\section{Master equations of the coarse-grained model}

In our coarse-grained approach, the structure is mapped into an effective 1D lattice 
by concentrating the occupation of the whole level onto one point $i\in\{0, 1, \ldots, n\}$, 
where $n$ is the depth of the tree. The detection point is located at the root and 
treated as an absorbing boundary. Thus, tracers that reach the root are not allowed 
to return to the tree. Let $E_i(t)$ denote the probability that a tracer particle 
enters the tree on level $i$ at time $t$. It is given by $E_i(t) = \frac{1}{t_e}\left(1 
-\frac{1}{t_e}\right)^t \frac{2^{i-1}}{2^n-1}$ for spontaneous uniform entering a 
binary tree used in the main text. Furthermore, let $q$ denote the hopping probability 
that a tracer particle moves during a timestep, $p$ denote the bias probability that 
a step is taken towards the root and $t_d$ the mean activation time for spontaneous 
signal emission after reaching the target. Then, the time evolution of the probability 
distribution $P_{i}(t)$ of a tracer being at depth level $i$ at time $t$ is governed 
by \cite{Jose18,Shaebani18}:
\begin{equation}
\left\{
\begin{aligned}
P_{0}(t +1)\,\,\,\,\,\, &= (1 -\nicefrac 1 {t_d}) P_{0}(t) + qp P_{1}(t), \\
P_{1}(t +1)\,\,\,\,\,\, &= (1 -q) P_1(t) +qp P_2(t) +E_1(t), \\
\vdots \\
P_{i}(t +1)\,\,\,\,\,\, &= q(1 -p) P_{i-1}(t) +(1 -q) P_{i}(t) +qp P_{i+1}(t) +E_i(t),\\
\vdots \\
P_{n-1}(t +1) &= q(1 -p) P_{n -2}(t) +(1 -q) P_{n-1}(t) +q P_{n}(t) +E_{n-1}(t),\\
P_{n}(t +1)\,\,\,\,\,\, &= q(1 -p) P_{n -1}(t) +(1 -q) P_{n}(t) +E_n(t).
\end{aligned}
\right.
\label{Eq:MasterEqs}
\end{equation}
These equations need to be solved for an initially empty tree, i.e., $P_i(0) \,{=}\, 0$ 
for all $i\,{\in}\,\{0,1,\ldots,n\}$. The signal is then given by $I(t+1) = \frac{1}{t_d} 
P_0(t)$.
\smallskip\smallskip\smallskip

In case the signal emission is not spontaneous but the emission times are distributed 
according to some distribution $\mathcal D(t)$, the signal can be expressed in terms 
of the first passage time $f(t)$ by the convolution
\begin{equation}
I(t) = \sum_{\tau=0}^t f(\tau)\mathcal D(t-\tau).
\end{equation}
Indeed, $f(t)$ is related to $P_1(t)$ via $f(t+1) = qpP_1(t)$ \cite{Jose18}. Since 
$P_0(t)$ does not feed back into the occupation probabilities of higher levels in 
Eqs.\,(\ref{Eq:MasterEqs}), it can be dropped in this scenario.

\section{Supplementary figures}

In this section, two additional figures illustrating the influence of the entering and 
emission times, $t_e$ and $t_d$, on the mapping between the model and shape parameters 
are presented.

\begin{figure}
\begin{center}
\includegraphics[width=0.9\textwidth]{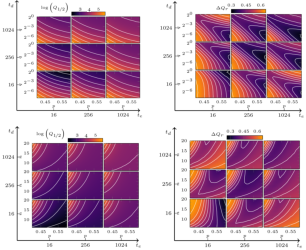} \\
\end{center}
Suppl.\,Fig.\,S2: Heat maps of the logarithm of the median $\log\!_{_{10}}\!
(Q_{\frac12}\!)$ (left) and the relative interquartile range $\Delta 
Q_r$ (right) in the $(q,p)$ plane (top) and the $(n,p)$ plane (bottom). 
Each single panel corresponds to a different set of mean entering time 
$t_e$ and mean emission time $t_d$ (presented by the outer axes). The 
same color coding is used for all heat maps in the left or right panels. 
The isolines of constant $\log\!_{_{10}}\!(Q_{\frac12}\!)$ or $\Delta 
Q_r$ are equidistantly chosen over the displayed ranges. Other parameter 
values:\ (top) $n\,{=}\,10$, (bottom) $q\,{=}\,0.5$.
\end{figure}

\begin{figure}
\begin{center}
\includegraphics[width=0.9\textwidth]{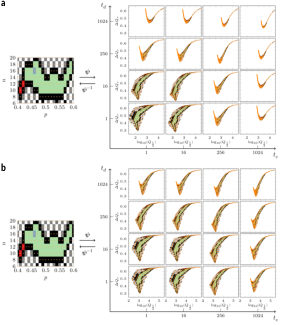} \\
\end{center}
Suppl.\,Fig.\,S3: Mapping of $(n,p)$ parameters to $\left(\logdec(\median),\dQr
\right)$ of the signal intensity, and vice versa. The mapping is presented for 
different mean entering $t_e$ and emission $t_d$ times and for (a) $q=2^{-1}$ 
or (b) $q=2^{-4}$. For every marked point in the model parameter domain (yellow 
crosses), the corresponding location in the shape parameter domain is extracted 
numerically. The neighborhood around each pair of connected points are painted 
with the same color in both domains for clarity.
\end{figure}

\end{widetext}

\end{document}